\documentclass[numberedappendix, onecolumn]{aa}

\usepackage{graphicx}
\usepackage{dcolumn}
\usepackage{bm}
\usepackage{mathrsfs}
\usepackage{amssymb}
\usepackage{amsmath}
\usepackage{subfigure}
\usepackage{booktabs}
\usepackage[square]{natbib}

\bibpunct{(}{)}{;}{a}{}{,}

\def\kte{kT_{\rm e}}

\def\gax {\ifmmode{_>\atop^{\sim}}\else{${_>\atop^{\sim}}$}\fi}

\def\cm2{cm$^{-2}$}
\def\s1{s$^{-1}$}
\def\mdot{$\dot{M}$}

\def\be{\begin{equation}}
\def\ee{\end{equation}}

\newcommand{\rf}[1]{(\ref{#1})}

\begin{document}
\title{Comptonization in ultra-strong magnetic fields: numerical solution of the radiative transfer problem}
\author{C. Ceccobello\inst{1,2}, R. Farinelli\inst{1,3} \& L. Titarchuk\inst{1,4}}
\offprints{C.Ceccobello, chiara.ceccobello@gmail.com}
\institute{
 Dipartimento di Fisica e Scienze della Terra, Universit\'a di Ferrara, via Saragat 1, 44122 Ferrara, Italy
\and INFN, Sezione di Ferrara, via Saragat 1, 44122, Ferrara, Italy
\and ISDC Data Center for Astrophysics, Universit\'e de Gen\`eve, chemin d'\'Ecogia 16, 1290 Versoix, Switzerland
\and NASA-GSFC, Greenbelt, MD 20771, USA\\}
\abstract{We consider the radiative transfer problem in a plane-parallel slab of thermal electrons in the presence of
an ultra-strong magnetic field ($B\gtrsim B_c\approx4.4\times10^{13}$ G). Under such conditions, the magnetic field behaves as a birefringent medium for the propagating photons, and  the electromagnetic radiation is split into two polarization modes, ordinary and extraordinary,
having different cross-sections. When the optical depth of the slab is large, the ordinary-mode photons are strongly Comptonized and the photon field is dominated by an isotropic component.}
{The radiative transfer problem in strong magnetic fields presents many mathematical issues and analytical or numerical solutions can be obtained only under some given approximations. We investigate this problem both from the analytical and numerical point of view, providing a test of the previous analytical estimates and extending these results introducing numerical techniques.}
{We consider here the case of low temperature blackbody photons propagating in a sub-relativistic temperature plasma, which allows us to deal with a semi Fokker-Planck approximation of the radiative transfer equation. The problem can be treated then with the variable separation method, and we use a numerical technique for finding solutions of the eigenvalue problem in the case of singular kernel of the space operator. The singularity of the space kernel is the result of the strong angular dependence of the electron cross-section in the presence of a strong magnetic field.}
{We report the numerical solution obtained for eigenvalues and eigenfunctions of the space operator, and the emerging Comptonization spectrum of the ordinary-mode photons for any eigenvalue of the space equation and for energies significantly less than the cyclotron energy, which is of the order of MeV for the intensity of the magnetic field here considered.}
{We derived the specific intensity of the ordinary photons, under the approximation of large angle and large optical depth. These assumptions allow the equation to be treated using a diffusion-like approximation.}


\authorrunning{Ceccobello et al. }
\titlerunning{Comptonization in ultra-strong magnetic fields}

\maketitle
\section{Introduction}
\label{intro}

Regardless on the astrophysical system considered, the presence of a magnetic field always plays a fundamental role in defining the characteristics and the dynamics of the ongoing phenomena. Only recently though, researchers are making steps forward in including magnetic fields in their pictures of several astrophysical systems, like, for instance, in the launching, acceleration and collimation of relativistic jets in Gamma-Ray Burst (e.g. \citealp{BlandfordZnajek:1977}) or in the inflow/outflow magnetically-channelled accretion process onto black holes (e.g. \citealp{Polko:2013}) and neutron stars (e.g. \citealp{ThompsonDuncan:1995}).\\
The increasing interest has stimulated dramatically the development of entire areas of research such as the magnetohydrodynamics (MHD) to understand how the plasma interacts with the magnetic field (see i.e. \citealp{Lithwick:2001,DeVilliers:2003,Proga:2003}). 
One of the most puzzling topic related to interaction between magnetic field and plasma is the physics of radiative transfer. Unfortunately an extensive and combined treatment of MHD and radiative transfer is still an open issue, since large numerical simulations are required for both fields. \\
To reduce the computational time, one possibility is finding a satisfactory semi-analytical treatment for radiative transfer within a set of reasonable \emph{a posteriori} assumptions made on the basis of observational constraints.\\
Analytical or numerical models with respect to a Montecarlo approach have the advantage of being much faster and, despite the significantly high degree of approximation, provide an insight to actual spectra based on the proper physical properties. \\
The main feature that emerges is that the problem is twofold, since the optical properties of the plasma are deeply affected by the presence of the magnetic field that modifies the interaction between photons and electrons through vacuum polarization. In particular, when a strong external magnetic field is present, higher order quantum effects in the computation of the vacuum polarization tensor of a photon within one-loop of a ``dressed'' fermion becomes important since the strong magnetic field compensates the smallness of the coupling constant giving rise to non-linear interactions among photons, e.g. fermion-antifermion pair creation, photon splitting \citep{HattoriItakura:2013}.\\
In presence of a plasma with a non-zero density gradient, the normal modes of photon propagation change from being circularly polarized at high electron densities to being mostly linearly polarized at low densities. The change in the polarization modes is accompanied also by a change in the opacities of the resulting normal modes due to vacuum polarization giving rise to significant resonances in the radiation-matter interaction. 
However, for sufficiently high magnetic fields ($B\gtrsim 10^{14}-10^{15}$ G) and plasma densities $n_e \gtrsim 10^{25}$ cm$^{-3}$ and for propagating photons of few keV, both modes are thermalized due to the high number of interactions with the plasma. As a result, including the off-diagonal vacuum polarization contributions to the dielectric tensor does not affect the results of the transfer calculations \citep{Ozel:2002}.\\
\noindent
Within the regime described above, when photons enter a magnetized plasma, they split into two polarization modes, depending on the orientation of their electric field with respect to $\hat k \wedge \overrightarrow B$, where $\hat k$ is the momentum of the photon and $\overrightarrow{B}$ is the external magnetic field. If $\overrightarrow E \parallel \hat k \wedge \overrightarrow B$ the polarization is called \emph{ordinary}, if instead $\overrightarrow E \bot \hat k \wedge \overrightarrow B$, the polarization is called \emph{extraordinary}.\\
The two polarization modes show different cross-sections for the interaction with matter, therefore a complete solution of the problem involves finding the emerging spectra of both populations of photons. However even deriving the exact cross sections for the radiation-matter interaction in presence of strong magnetic fields appeared to be quite demanding especially when the external magnetic field reaches or exceeds the critical value $B_{\rm c}=4.413\times10^{13}$ G so that $h\nu_g\approx m_ec^2$, where $\nu_g$ is the cyclotron frequency which is given by the relation
\be
h\nu_g=\hbar\frac{e B}{m_e c}=11.57\,B_{12}\,\rm{keV}, \label{hnu_g}
\ee
where $B_{12}=B/(10^{12} \rm G)$. 
The full scattering cross sections were derived in several steps across the years (\citealp{Ventura:1979,Herold:1979,Melrose:1983,Daugherty:1986,Harding:1991}). \\
\cite{Meszaros:1984} discussed the propagation of polarized radiation in strong magnetic field in terms of wave equation in the Fourier space considering quantum electrodynamics (QED) effects, polarization and mode ellipticity for Thompson, bremsstrahlung and Compton scattering processes. The dependence on angles and frequency of the cross-sections arose several problems, so the author proposed time-dependent, averaging and approximation techniques as a way out to this numerical empasse.\\
\cite{Pavlov:1989} presented a comparison of the Comptonization process in magnetized and unmagnetized optically thick plasmas. Using a simplified analytical treatment and including stimulated scattering, obtaining spectra at energies lower than the plasma. temperature. \\
Later on \cite{L86} and \cite{L88I,L88II}, hereafter collectively L88, proposed an extensive analysis of inverse Compton scattering of soft photons in a strongly magnetized non-relativistic medium. The author considered approximated cross-sections which maintain both the angular and energy dependencies. Using a form of the Fokker-Plank approximation which allows to maintain the integro-differential nature of the radiative transfer equation (RTE), he managed to find emerging spectra of ordinary photons.\\  
\cite{LyutikovGavriil:2006} proposed a semi-analytical treatment of
the resonant Compton scattering (RCS) process in a plane-parallel slab (axis-symmetric) configuration. Their model consists in the study of Thomson scattering process of a blackbody (BB) spectrum in a static plasma filled with an electron population at constant temperature and density. Despite these crude approximations, this model was successfully applied by \cite{Rea:2007a,Rea:2007b,Rea:2008} to magnetar spectra below 10 keV.\\
Most recently \cite{NTZ08}, hereafter NTZ08, developed a 3D Monte Carlo twisted magnetospheric model for the RCS assuming that isotropic and unpolarized BB photons are emitted at the neutron star surface. \\
NTZ08 applied their model to a sample of Anomalous X-ray Pulsars (AXPs) and Soft Gamma Repeaters (SGRs) observed by \textit{XMM} and \textit{INTEGRAL} in order to have the as large as possible energy coverage.
They found that the \textit{XMM} spectra alone are well described by their model, while for sources observed also with \textit{INTEGRAL} (1RXS J1708-4009, 1E1841-045, SRG 1900+14) the observed spectra require an additional power law component with photon index $\Gamma \sim$ 1-2.\\
The Montecarlo approach used by NTZ08 has the advantage of allowing to treat the problem with enough high degree of accuracy, considering both the scattering process (QED effects) and the geometry of the problem (multipolar components of the magnetic field, non-uniform electron temperature and bulk velocity). \\
A detailed analytical and numerical treatment of the radiative transfer problem in the presence of a relatively strong magnetic field for plasma subjected to inward bulk motion was developed by \cite{BeckerWolff:2007}, hereafter BW07. BW07 used a Fokker-Planck approximation for the RTE (see also \citealp{BlandfordPayne:1981}) for the particular case of cylindrical, plane-parallel geometry, by including a modified angular-integrated form of the Thomson cross-section for taking into account photon diffusion in space and energy over the plasma configuration. However, they considered the transfer equation for the zero-moment of the photon occupation number, which thus did not take into account the possible strong anisotropy in the radiation field specific intensity.\\
Recently \cite{Farinelli:2012a}, hereafter F12, developed a numerical code aimed at performing spectral fitting analysis of magnetized sources with $B\gtrsim  10^{12}$ G, like X-ray pulsars and supergiant fast x-ray transients (SFXT). They adapted a relaxation method to seek for solutions of the radiative transfer problem in diffusion approximation. We refer the reader to F12 for a detailed discussion of the numerical code. 
The authors are applying successfully the model to several SFXTs, as reported in \cite{Farinelli:2012b}.\\
Here, we consider the same configuration of L88, i.e. a plane-parallel slab of thermal plasma consisting in non-relativistic electrons
having Thomson optical depth $\tau_0$ and temperature $kT_{e}$. A uniform magnetic field is oriented along the normal of the slab and has a strength of order of $10^{14}$ G or higher.
A thermal bath of photons distributed as the Planck's function with temperature $kT_{bb}$ penetrates the slab from the bottom, acquiring a polarization which can be either ordinary (``O'' subscript) or extraordinary (``X'' subscript). \\
Vacuum polarization contributions and resonant scattering can be neglected considering photon energies below $10$ keV, magnetic fields of the order of $10~B_c$ or larger and electron densities $\gtrsim 10^{25}$ cm$^{-3}$.\\
In this paper, we reproduce the analytical results obtained in L88 and we extend them introducing numerical techniques to solve the RTE of the ordinary photons. \\
We compare our numerical results with the case of zero-magnetic-field discussing the main differences between the two cases.\\
In Section \ref{analytic} we describe the system and the first assumptions we made following L88 in order to get a RTE for the ordinary photons. In Section \ref{problem_solution} we define the range of angles and optical depth within which we can apply the Fourier method to the RTE. In Section \ref{ASmethod} we discuss in details how to solve the equation for the space variable, i.e. the optical depth, while in Section \ref{Green} is reported how to obtain the solution of the energy equation. In Section \ref{intensity_pol} we describe how calculate the photon angular distribution inside the slab and the specific intensity.
Section \ref{results} is dedicated to the discussion of the results we obtained from the numerical code and the comparison with the unmagnetized case. Finally, in Section \ref{Conclusions} we draw our conclusions.

\section{Problem settings}
\label{analytic}
  
We consider a plane-parallel slab of non-relativistic electrons dipped into a magnetic field $B>4.4\times 10^{13}$ G, which is uniform and directed along the normal of the slab. \\
Since charged particles follow helicoidal trajectories along the magnetic field lines with a gyro-radius which is inversely proportional to the magnetic field strength, a one-dimensional Maxwell-Boltzmann describes quite accurately the distribution of the electrons. \\
We consider non-resonant scattering, therefore photon energies well below the cyclotron energy
($h\nu\ll h\nu_g$), so that the photons have two normal polarization modes \citep*{Lai:2001}.\\
During the scattering process, photons are allowed either to maintain their polarization mode or flip it.
The approximated magnetic Thomson scattering differential cross-sections for the interaction with the plasma are:
\begin{align}
d\sigma_{\rm O\rightarrow\rm O}(\mu, \mu') & = \frac{3}{4}\sigma_{\rm T}\left[ (1-\mu^{2})(1-\mu'^{2})+\frac{1}{2}\left(\frac{\nu}{\nu_{\rm g}} \right)^{2}\mu^2\mu'^2\right]d\mu' \label{crosssec2}\\
d\sigma_{\rm X\rightarrow\rm X}(\mu, \mu') & = \frac{3}{8}\sigma_{\rm T}\left(\frac{\nu}{\nu_{\rm g}} \right)^{2}d\mu' \label{crosssec1}\\
d\sigma_{\rm O\rightarrow\rm X}(\mu, \mu') & = \frac{3}{8}\sigma_{T}\left(\frac{\nu}{\nu_{\rm g}} \right)^{2}\mu^{2}~ d\mu' \label{crosssec21}\\
d\sigma_{\rm X\rightarrow\rm O}(\mu, \mu') & = \frac{3}{8}\sigma_{T}\left(\frac{\nu}{\nu_{\rm g}} \right)^{2}\mu'^{2}~ d\mu'\label{crosssec12}
\end{align}
where $\mu=\cos\psi$ and $\mu'=\cos\psi'$ are the cosines of angles between the direction of the magnetic field and the direction of photons before and after the scattering, respectively, while $\sigma_T$ is the Thomson cross-section.\\
It is worth noticing that only the first term on the right hand side of equation ~(\ref{crosssec2}) does not depend on frequency and on magnetic field through $\nu_g$. under the assumption of magnetic fields larger than the critical value $B_c$ and non resonant scattering, $\nu \ll \nu_g$. The first term in  equation \rf{crosssec2} dominates over all the other ones, except for photons propagating parallel or anti-parallel to the magnetic field lines.\\
In this regime, the Comptonization process affects the emerging spectrum of the ordinary photons, leaving almost unchanged the spectrum of the extraordinary photons. We concentrate then on the study of the radiative transfer process of the ordinary photons, although the spectral feature related to the extraordinary photons can provide an important contribution to the total emerging spectrum.
The extraordinary photon component that will be considered here originates from the fraction of ordinary photons that have changed their polarization during the scattering process according to the cross-section \rf{crosssec21}. \\

\section{Solution of RTE in a Magnetized Medium}
\label{problem_solution}

The homogeneous integro-differential form of the RTE for the ordinary mode photons, neglecting induced processes and considering inverse Compton as the leading process, is given by
\begin{align}
 \mu\frac{\partial}{\partial r} n_{\rm O}(\mu,\nu, \vec{r})=& \int dp\, d\sigma_{\rm O\rightarrow \rm O} \{n_{\rm O}(\mu',\nu', \vec{r})N_e(p') \nonumber\\
& - n_{\rm O}(\mu,\nu, \vec{r}) N_e(p) \} \nonumber\\
&-\sigma_{\rm O\rightarrow \rm X}\bar{N}_{e}\,n_{\rm O}(\mu,\nu, \vec{r}) \label{lyubeqn}
\end{align}
where $n_O(\mu, \nu,\vec r)$ is the ordinary photon occupation number, $N_e(p)$ is the electron distribution function, that we assume to be the one-dimensional Maxwellian and $\bar{N}_{e}$ is the electron density. For the sake of simplicity, hereafter we will use the simplified notations $\sigma_{\rm{OO}}$ and $\sigma_{\rm{OX}}$ for the cross sections \rf{crosssec2} and \rf{crosssec21}. \\
In order to find an approximate solution of the integro-differential equation (\ref{lyubeqn}), we expand both the occupation number and the Maxwellian electron distribution in Taylor series up to second order in $\Delta \nu$ and $\Delta \epsilon$, respectively.\\
Note that, where an external magnetic field is present, even after the Taylor expansion, equation \rf{lyubeqn} does not reduce to a purely differential equation both for the space and the energy variable (see eq.~8 of L88), like it does in the general case of zero magnetic field (\citealp{Rybicki:1979}), hence we are still dealing with an integro-differential equation.\\
Following the arguments reported in L88, it worth noticing that the Comptonization parameter, which in strong magnetic fields is $y_{\rm{NR}}\approx(2/15)(4kT_{\rm e}/m_ec^2)\tau^2$, is smaller than in the unmagnetized case by a factor of $\sim 0.13$. Thus, we need to assume that our system is optically thick ($\tau\gg 1$), so that the average number of scatterings may be large enough to make the Comptonization process effective.\\
Due to the presence of the magnetic field, the cross section \rf{crosssec1} approach zero for angles $\psi < \tau^{-1}$, therefore photons travelling at such small angles with respect to the field are escaping freely from the plasma (\citealp{Lyubarskii:1982}). \\
Clearly, only the photons which move at sufficiently large angles to the field undergo enough scatterings to be effectively Comptonized, i.e. the emerging spectrum will be significantly modified with respect to the seed spectrum.
At such large angles the optical depth $\tau$ is as large enough to ensure that Comptonized photons diffuse almost isotropically. Therefore, under the conditions of large angles and large optical depths, we neglect the anisotropic part $\delta n$ of the occupation number $n=n_{iso}+\delta n\approx n_{iso}$.
The resulting equation may be handled with the separation of variable method.\\
We define the dimensionless energy $x\equiv h\nu/kT_{\rm e}$ and we replace the space variable $r$ with the Thompson optical depth for electron scattering $\tau= N_e \sigma_T r$. We seek a solution of the form $n(x, \tau)= s(\tau)Z(x)$. Substituting it into equation (\ref{lyubeqn}) and, introducing a source distribution accounting for the seed ordinary photons $\mathcal S(x)$, we obtain the following two independent equations 
\begin{equation*}
\left\{
\begin{aligned}
 \mathcal L_x \mathcal Z(x)&=\lambda~ \mathcal Z(x),\nonumber\\
 \mathcal L_\tau s(\tau)&=\lambda~ s(\tau)
\end{aligned}
\right.
\end{equation*}
or, more explicitly
\begin{align}
&\frac{1}{x^2}\frac{d}{dx} x^{4}\left(\frac{d}{dx}\mathcal Z(x)+ \mathcal Z(x)\right)-(lx^2+\gamma)\mathcal Z(x)=\mathcal S(x), \label{energyeqn}\\
&\left(1-\frac{3}{4}\lambda\right) s(\tau)=\int_{-\tau_{0}}^{\tau_{0}}d \tau' K(|\tau-\tau'|) s(\tau'),\label{spatialeqn}
\end{align} 
where we have defined the quantities 
\begin{equation}
 \gamma=\frac{15}{2}\frac{m_ec^{2}}{kT_{\rm e}}\lambda \qquad \textrm{and} \qquad l=\frac{15}{8}\frac{m_ec^{2}}{kT_{\rm e}}\frac{1}{{x_{\rm g}}^{2}}\label{gamma_lambda}
\end{equation}
and the kernel 
\begin{equation}
 K(|\tau-\tau'|)= \frac{3}{4}\int_0^1 \frac{(1-\mu^2)^2}{\mu}e^{-\frac{(1-\mu^2)}{\mu}|\tau-\tau'|}d\mu\,.\label{kerneltheta}
\end{equation}
The quantity $\lambda$ in equation (\ref{spatialeqn}) is related to the eigenvalue of the space operator $\mathcal L_\tau$ for the eigenfunctions $s(\tau)$.\\
Equation (\ref{energyeqn}) is similar to the standard Kompaneets diffusion equation, with no induced processes, however the term depending on the $l$-parameter on the left hand side contains the magnetic field dependence, via $x_{\rm{g}}$, which is the dimensionless cyclotron energy of the electron, namely $x_g\equiv h\nu_g/kT_{\rm e}$.\\
Although the additional ``magnetic'' term in equation (\ref{energyeqn}) affects the dynamics of the Comptonization process, the equation maintains approximately the same mathematical structure, so it can be solved with the standard Green's function method (see Sect.~\ref{Green}). \\
Equation (\ref{spatialeqn}) is instead an homogeneous Fredholm equation of the second kind with the logarithmically singular kernel (\ref{kerneltheta}). It is worth noticing that in the unmagnetized case the space equation obtained from the variable separation is described by a differential operator (\citealp{Rybicki:1979}). \\
The solution of equation (\ref{spatialeqn}) is not straightforward, since the standard integration techniques cannot handle with a kernel singularity, even if it is moderate. We adopt an algorithm suggested by \cite{Atkinson:2008}, described in Sect.~\ref{ASmethod}, which is thought specifically for kernels with a ``quasi''-smooth behaviour. \\
The overall solution of the radiative transfer problem for the isotropic part of the ordinary photon occupation number described by the system \rf{energyeqn}-\rf{spatialeqn} should be found in the form (see also \citealp{TMK:1997}, hereafter TMK97) 
\begin{equation}
n(x,\tau)= \sum_{k=1}^{\infty}n_k(x,\tau)=\sum_{k=1}^{\infty}c_{k} s_{k}(\tau)\mathcal Z_{k}(x)\,, \label{solution}
\end{equation} 
where $s_{k}(\tau)$ is the $kth$-eigenfunction of equation (\ref{spatialeqn}) and $\mathcal Z_{k}(x)$ is the solution of equation (\ref{energyeqn}) for the $kth$-eigenvalue. The coefficients $c_k$ are the Fourier coefficients of the series, i.e. the projections of each eigenfunction over a properly chosen spatial photon distribution $f(\tau)=e^{-\tau/2\tau_0}$.

\section{Solution of the Spatial Problem}
\label{ASmethod}
The space problem of RTE (see eq.~\ref{spatialeqn}) is an homogeneous Fredholm integral equation of the second kind, i.e. an eigenvalue problem, namely
\begin{equation}
L_\tau s(\tau)\equiv\int_{-\tau_0}^{\tau_0}K(|\tau-\tau'|)s(\tau')d\tau'=\sigma s(\tau)	
\label{Loperator}
\end{equation} 
where $\sigma=\left(1-\frac{3}{4}\lambda\right)$ and the kernel is reported in equation (\ref{kerneltheta}).
Even if the kernel has a logarithmic singularity along its diagonal ($\tau=\tau'$), it is possible to demonstrate that the integral operator $L_\tau$ maintains the property of compactness.
Indeed, performing the integration over $\tau'$, we obtain an analytical expression for the integrand function
\begin{align}
 \int_{-\tau_0}^{\tau_0}K(|\tau-\tau'|)d\tau'= &\frac{3}{4}\int_0^1 (1-\mu^2)\left[2-e^{-\frac{(1-\mu^2)}{\mu}(\tau_0+\tau)}\right.\nonumber\\
 &\left.-e^{-\frac{(1-\mu^2)}{\mu}(\tau_0-\tau)}\right]d\mu\,. 
\end{align}
The integrand is smooth and the integral is finite, so the integral operator $L_\tau$ is a compact operator, thus it is bounded and it has a complete set of eigenvalues and eigenfunctions (see \citealp{Atkinson:1967}). 

\subsection{Numerical Treatment of the Singularity}

\begin{figure}[t]
 \centering
  \includegraphics[scale=.7]{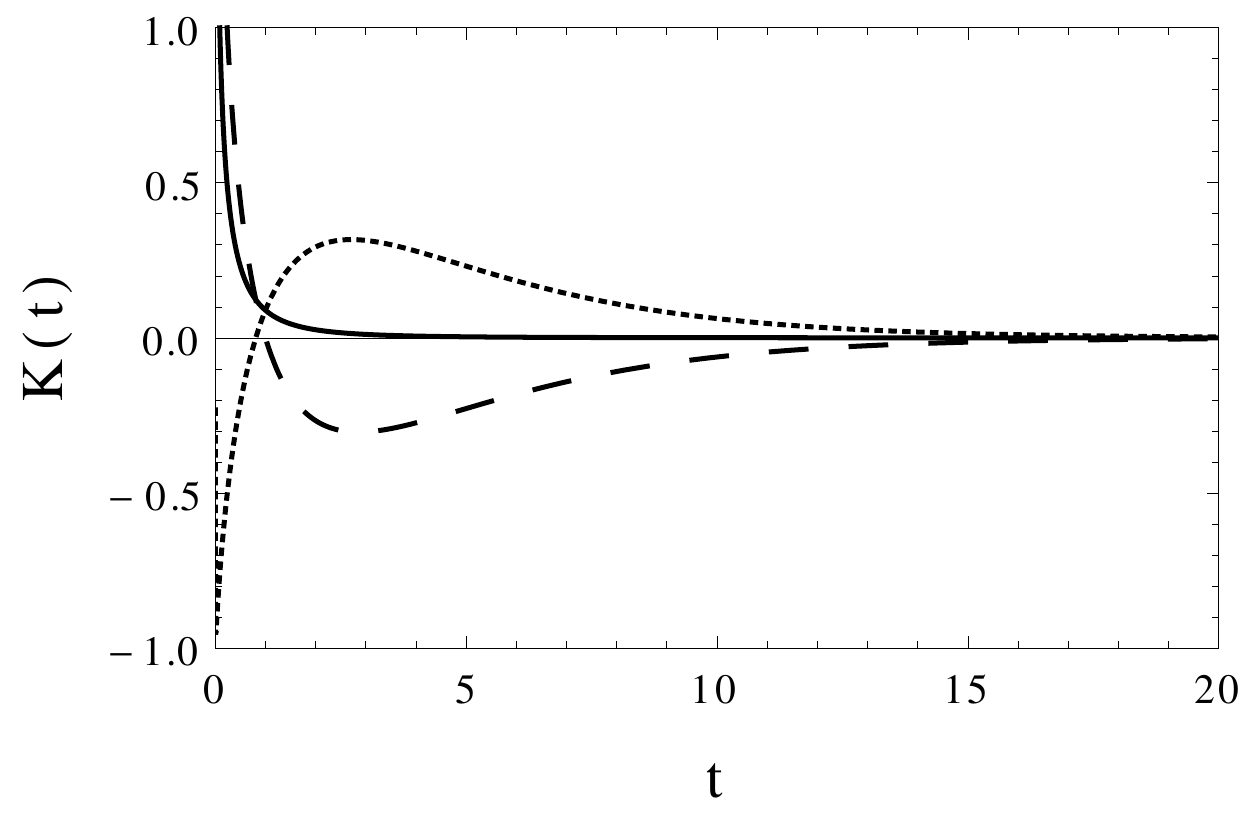}
  \caption{Numerical representation of the kernel (\ref{logkernel}) with $t=|\tau-\tau'|$. Dotted line: the singular logarithmic part of the kernel. Dashed line: the smooth polynomial term of the kernel. Solid line: the total kernel, sum of the two contributions.} 
  \label{fig:kernel} 
\end{figure}

Although, in principle, a logarithmic singularity is integrable, we may have numerical issues in mapping the solution in the regions of integration near the boundaries. However, the kernel can be transformed analytically before the numerical integration of equation \rf{spatialeqn} takes place. In particular, defining the variable such as $t=|\tau-\tau'|\rightarrow0$ and expanding the kernel around $t$, we find 
\begin{align}
K(t) & =\frac{3}{4}\int_0^1 \frac{(1-\mu^2)^2}{\mu}e^{\mu t}e^{-\frac{t}{\mu}}d\mu \nonumber\\
& \approx \frac{3}{4}\int_0^1 \frac{(1-\mu^2)^2}{\mu}\left[1+\mu t+\frac{(\mu t)^2}{2!}+\dots\right]e^{-\frac{t}{\mu}}d\mu\,. \label{kernelexp}
\end{align}
With a further change of variable $\mu=1/y$, it is possible to write the kernel as a sum of exponential integrals  \citep*{AbramovitzStegun:1970}, that for $Re(t)>0$, are defined as 
\be
E_n(t)=\int_1^\infty \frac{e^{-yt}}{y^n}dy\,. 
\ee
The exponential integrals $E_n(t)$, if $|\arg\,t|<\pi$, can be written also in the form \citep*{Bleistein:1986}
\be
E_n(t)=\frac{(-t)^{n-1}}{(n-1)}\left[-\log t+\psi(n)\right]-\sum_{\substack{m=0\\m\neq n-1}}^{\infty}\frac{(-t)^m}{(m-n+1)m!} \label{expint_series}
\ee
where
\be
\psi(1)=-\gamma,\qquad \psi(n)=-\gamma+\sum_{m=1}^{n-1}\frac{1}{m}\qquad \qquad(n>1)
\ee
and $\gamma=0.57721...$ is Euler's constant. The series of exponential integrals (\ref{expint_series}) can be written more concisely as
\be
E_n(t)=-P(t)\log t+Q(t), 
\ee
where $P(t)$ is a polynomial and $Q(t)$ is a series around $t$, therefore the kernel (\ref{kerneltheta}) takes the form 
\begin{equation}
K(|\tau-\tau'|)=\frac{3}{4}\left[-\log(|\tau-\tau'|)+Q(|\tau-\tau'|)\right],\label{logkernel}
\end{equation}
in which $P(|\tau-\tau'|)=1$ \citep*{AbramovitzStegun:1970}. In this explicit form, the logarithmic singularity has been separated from the regular part $Q(|\tau-\tau'|)$. \\
Whereas the smooth part can be easily treated with any of the standard quadrature rules, the integration of the logarithmic term requires more attention when $t \rightarrow 0$ as shown in Fig.~\ref{fig:kernel} where we plot separately the behaviour of the two components of the kernel in equation \rf{logkernel} as a function of the variable $t=|\tau-\tau'|$, with $\tau_0=20$ and the total kernel.
Indeed, a direct integration over the logarithmic part of (\ref{logkernel}) is not straightforward to perform with the standard analytical and numerical integration techniques (e.g. \citealp{Morse:1953, Polyanin:2008}). 

\subsection{The Atkinson-Shampine Method}

We adopt the algorithm described by \cite{Atkinson:2008}, hereafter AS08. They present a numerical {\large M}{\small ATLAB} program, called \texttt{Fie}, which has been created with the purpose of providing a numerical code for the integration of Fredholm integral equations of the second kind on a interval that can be either finite $\left[a,b\right]$ or semi-infinite $\left[0,\infty\right)$. The authors considered not only kernels $K(s,t)$ that are smooth functions on $R=\left[a,b\right]\times\left[a,b\right]$, but also kernels having a modest singularity behaviour across the diagonal $s=t$, as long as they conserve compactness.\\
The integral equations that \texttt{Fie} is called to solve are 
\be
\lambda\, x(s)-\int_a^b\log|s-t|\,x(t)\,dt=f(s),\qquad a\leq s\leq b, \label{eqn:Fredlog}
\ee
and they may have solutions $x(s)$ which cannot be necessarily smooth at the boundaries of the integration interval. 
To account for the singular behaviour across the diagonal, AS08 uses a product Simpson's rule for the integration.\\
Hence, AS08 introduced a mesh of integration points $\{t_0, \dots,t_n\}$ which is \emph{graded} near the integration limits, $a\,\textrm{and}\,b$, where the behaviour of the solution can be critical. The index $n$ is always chosen to be divisible by 4, and sufficiently large, in order to guarantee the existence of a unique solution of the problem.\\
The solution is requested to satisfy the convergence criterion for $n\rightarrow\infty$
\be
||x-x_n||_\infty\leq c\,||\mathcal K\, x-\mathcal K_n\, x||_\infty \qquad \qquad (c>0), \label{error}
\ee
where we have defined the integral operators
\be
\mathcal K\,x=\int_a^b\log|s-t|\,x(t)\,dt\,,
\ee
and $\mathcal K_n$ is its approximated form, that we will describe later on.
Inequality \rf{error} holds if the separation between the mesh points is chosen properly. Roughly speaking, the grading of the mesh should be intensified near the critical points for the integration. In particular, we want that the error $||x-x_n||_\infty$ should be, at least, of order $\mathcal O(n^{-p})$ with $p=3$. 
The general integration scheme, suggested by AS08, says that for any triplet of points $\{t_{j-1},t_j,t_{j+1}\}$ with $j$ odd, the solution $x(t)$ is approximated with a piecewise quadratic interpolation function $\tilde x_j(t)$, so that the integral $\mathcal K\, x(t)$ becomes
\begin{align}
\mathcal K\, x(t)&\equiv\int_{a}^{b}\log|s-t|\,x(t)\,dt = \sum_{\substack{j=1,\\j\,\textrm{\footnotesize{odd}}}}^{n-1}\int_{t_{j-1}}^{t_{j+1}}\log|s-t|\,x(t)\,dt \nonumber\\
&\approx\sum_{\substack{j=1,\\j\,\textrm{\footnotesize{odd}}}}^{n-1}\int_{t_{j-1}}^{t_{j+1}}\log|s-t|\,\tilde{x}_{j}(t)\,dt\nonumber\\
& = \sum_{k=0}^{n}M_k(s)\,x(t_k)\equiv \mathcal K_n\, x(s),
\end{align}
where $M_k(s)$ are the \emph{weights} of the interpolating function over each subinterval. 
Thanks to the property of compactness of both terms of \rf{logkernel}, we use the same mesh to calculate the non-singular part of the kernel too and, thus, we define the total weight matrix $M_k(\tau)=M_k^S+M_k^{NS}$. \\
The solution of equation \rf{spatialeqn} is found solving the algebraic equation that for each eigenvalue $\sigma$ is the following
\begin{equation}
 \sigma x(s)=\sum_{k=0}^{n}M_k(s)x(t_k),\qquad a < s < b\,.
\end{equation}
It is worth noticing that the actual eigenvalues of \rf{spatialeqn}, $\lambda_k=\frac43\left(1-\sigma_k\right)$, are those that appears in the last term of the left hand side of equation \rf{energyeqn} in the parameter $\gamma$.

\section{Green's Function of the RTE Energy Operator}
\label{Green}

The energy operator \rf{energyeqn} has the form of a confluent hypergeometric equation and a source term which can be solved with the Green's function method (see e.g. TMK97 and BW07). After collecting terms, we obtain a more explicit form of equation \rf{energyeqn}, which is
\begin{equation}
x^{2}\frac{d^{2}\mathcal G}{dx^{2}}+\left(4x+x^{2}\right)\frac{d\mathcal G}{dx}+\left(4x-lx^{2}-\gamma \right)\mathcal G=\frac{\delta(x-x_{0})}{x^{3}},
 \label{startingpoint}
\end{equation} 
where the source term $\mathcal S(x)$ on the right hand side has been replaced by a delta function $\delta(x-x_0)/x^3$ representing a monochromatic source of injected photons.\\
The Green's function, solution of equation \rf{startingpoint} can be expressed in terms of the Witthaker functions as 
\be
{\mathcal G}(x, x_0)=\frac{e^{\frac{x_0}{2}\left(1-\sqrt{1+4l}\right)}}{x_0~\Gamma (2\alpha+4)} \times\nonumber\\
\ee
\begin{align}
& \left\{
\begin{aligned}
 &_1\mathcal F_1(\alpha,k,l,x){e^{-\frac{x}{2}\left(1+\sqrt{1+4l}\right)}}~\left(\frac{x } {x_0} \right)^{\alpha +3}I(x_0,\alpha,l), \, x \le x_0,\\
 &_1\mathcal F_1(\alpha,k,l,x_0){e^{-\frac{x_0}{2}\left(1+\sqrt{1+4l}\right)}}\left(\frac{x}{x_0} \right)^{-\alpha} I(x,\alpha,l), \, x \ge x_0 ,\label{eqn:Gfunc}
\end{aligned}
\right.
\end{align}
where we have defined
\begin{align}
& _1\mathcal F_1(\alpha,k,l,x)= _1F_1(\alpha+2-k,4+2\alpha,x\sqrt{1+4l})\nonumber\\
& _1\mathcal F_1(\alpha,k,l,x_0)= _1F_1(\alpha+2-k,4+2\alpha,x_0\sqrt{1+4l})\nonumber\\
& I(\alpha,x,l)=\int_0^\infty(x\sqrt{1+4l}+t)^{\alpha+1+\frac{2}{\sqrt{1+4l}}}t^{\alpha+1-\frac{2}{\sqrt{1+4l}}}e^{t}dt, \label{steepint}
\end{align}
and the slopes of the two power laws are determined by the spectral index
\begin{equation}
\alpha_k=-\frac{3}{2}+\sqrt{\frac{9}{4}+\frac{15}{2}\frac{m_ec^{2}}{kT_{\rm e}}\lambda_k},
\label{specindex}
\end{equation} 
which depends on the $k$-th eigenvalues of equation \rf{spatialeqn}.
The $k$-th spectrum can be finally expressed as 
\begin{equation}
\mathcal Z_{O,k}(x) =\int^{\infty}_0 \mathcal G_O(x,x_{0},\lambda_k) \mathcal S(x_0)\, dx_0. \label{flux2}
\end{equation}

\section{Angular Distribution and Specific Intensity}
\label{intensity_pol}
The solution for the photon occupation number of the system \rf{energyeqn}-\rf{spatialeqn} for a particular eigenvalue is
\begin{equation}
n_k(\nu, \tau)=s_k(\tau)\mathcal Z_k(\nu),
\end{equation} 
thus we can write the specific intensity $I=(2h \nu^{3}/c^{2})n$ as a series of terms which are products of two functions with separated dependencies on the variables $x$ and $\tau$. Therefore, apart from dimensional factors, the specific intensity is 
\begin{equation}
I(x,\mu, \tau)\approx \sum_{k=1}^{\infty}c_kJ_k(\mu, \tau)\mathcal Z_k(x), \label{sint}
\end{equation}  
where $\mathcal Z(x)$ is given by \rf{flux2}, in which we dropped the label ``$O$'' for the sake of clarity. On the other hand, the angular distribution is related with the eigenfunctions $s_k(\tau)$, solutions of the space problem \rf{spatialeqn}, as described in the following relation
\begin{equation}
J_k(\mu, \tau)= \left\{
\begin{aligned}
\,\,(1-\mu^{2})\int_{-\tau_{0}}^{\tau} e^{- \frac{(1-\mu^{2})}{\mu}(\tau- \tau')} s_k(\tau') \frac{d\tau'}{\mu}\,,& \, \mu>0,\\
-(1-\mu^{2}) \int_{\tau}^{\tau_0} e^{\,\frac{(1-\mu^{2})}{\mu}(\tau'- \tau)} s_k(\tau') \frac{d\tau'}{\mu}\,,& \, \mu<0.\\
\end{aligned}
\right.
\label{bc_ang}
\end{equation}
The coefficients $c_k$ of the expansion of the series \rf{solution} are obtained from the projection of the eigenfunctions over the spatial distribution of the source, i.e.
\be
c_k=\int_{-\tau_0}^{\tau_0}s_k(\tau')f(\tau')d\tau',
\ee
where $f(\tau)$ is a given spatial distribution of the seed photons over the bounded medium.\\
In addition to the specific intensity carried by ordinary photons, since in equation \rf{lyubeqn} we have included also the term which accounts for mode-switching from $O$ to $X$, we should calculate the contribution to the total specific intensity provided by such population of extraordinary photons.\\
The term accounting for the creation of extraordinary photons is determined by the cross section \rf{crosssec21}, that after the integration over $\psi'$ becomes angle-independent, thus the extraordinary photons originated via mode-switching $O\rightarrow X$, can be considered isotropically distributed, with a spectral shape (see L88)
\be
\mathcal Z_X(x)=\frac{1}{4}\left(\frac{x}{x_g}\right)^2\sum_{k=1}^n \mathcal Z_{O,k}(x)\int_{-\tau_0}^{\tau_0}s_k(\tau)d\tau, \label{sintE}
\ee 
where $\mathcal Z_{O,k}(x)$ as been calculated in \rf{flux2}.
The flux of extraordinary photons turns out to be quite small with respect to the sum of $\mathcal Z_{k,O}$ over all $k$ due to the inverse proportionality with the magnetic field ($x_g\sim B$) and the modulation brought by the eigenfunctions $s_k(\tau)$.

\section{Results}
\label{results}

In this Section we present the results in several steps to discuss separately the solutions of equations ~\rf{spatialeqn}-\rf{energyeqn} and compare them with similar results for the same system in absence of magnetic fields.

\subsection{Eigenvalues and eigenfunctions of the space operator}
\label{space}
\begin{figure}[t!]
 \centering
 \includegraphics[scale=.8]{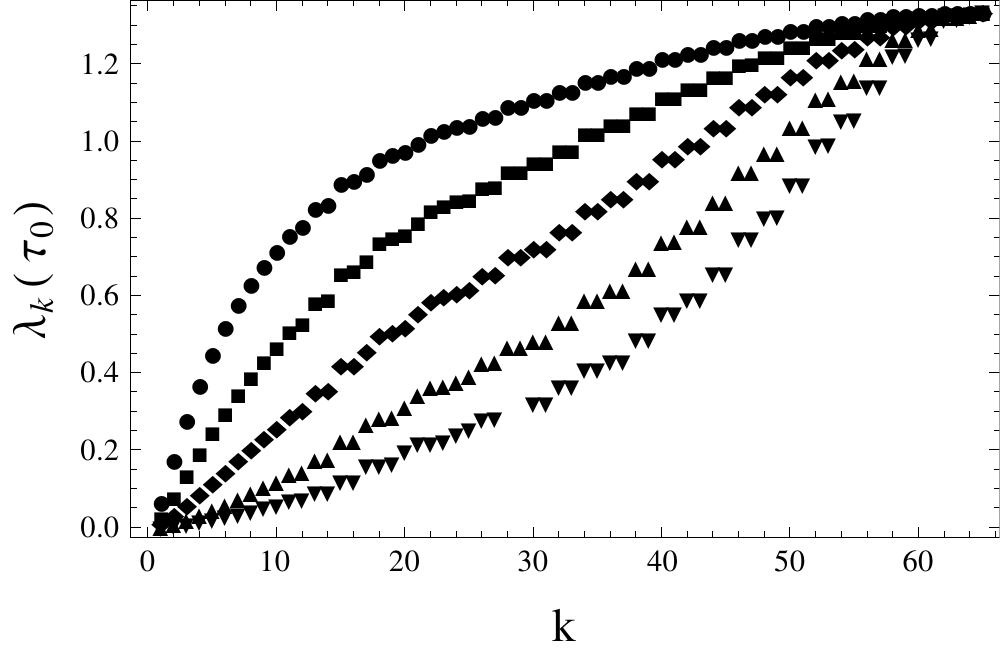}
 \caption{Five series of eigenvalues of equation \rf{spatialeqn} for optical depths $\tau_0=5,10,20,40,70$. Filled dots: 64 eigenvalues for $\tau_0=5$. Filled squares: same set of eigenvalues for $\tau_0=10$. Filled rhombuses: same set of eigenvalues for $\tau_0=20$. Filled up triangles: $\tau_0=40$. Filled down triangles: $\tau_0=70$.}
\label{fig:eigenv}
\end{figure}

The algorithm described by AS08, in principle, allows to find all the terms of the infinite series of eigenvalues and eigenvectors of equation \rf{spatialeqn}. Nonetheless, the limitation comes from the numerical accuracy.\\
In Fig.~\ref{fig:eigenv} we show five sets of $64$ eigenvalues for increasing maximum optical depth $\tau_0$ (from filled circles to filled down triangles).
\noindent
The number of eigenvalues that we extrapolate from the algorithm is equivalent to the number of points of the grid $n$ or less. The limitation that we encounter is numerical and it is related with $n$. \\
From Fig.~\ref{fig:eigenv}, we notice that, above a certain $k$, the eigenvalues start to be indistinguishable, showing also features of numerical degeneracy and, as the optical depth increases, the problem migrates to lower $k$. Letting $n$ increase, we are able to push the degeneracy to higher orders at the expenses of computational time.
With $n=64$, we consider eigenvalues up to $k=10$ in order to avoid this numerical issue.   \\
The truncation of the series at $k=10$ however doesn't affect our analysis since, as we will show later on in this Section, the high energy (Comptonized) part of the spectrum is mainly determined by the first term $k=1$, while higher orders contributes to the soft peak. 

\begin{figure}[t!]
 \centering
 \includegraphics[scale=.85]{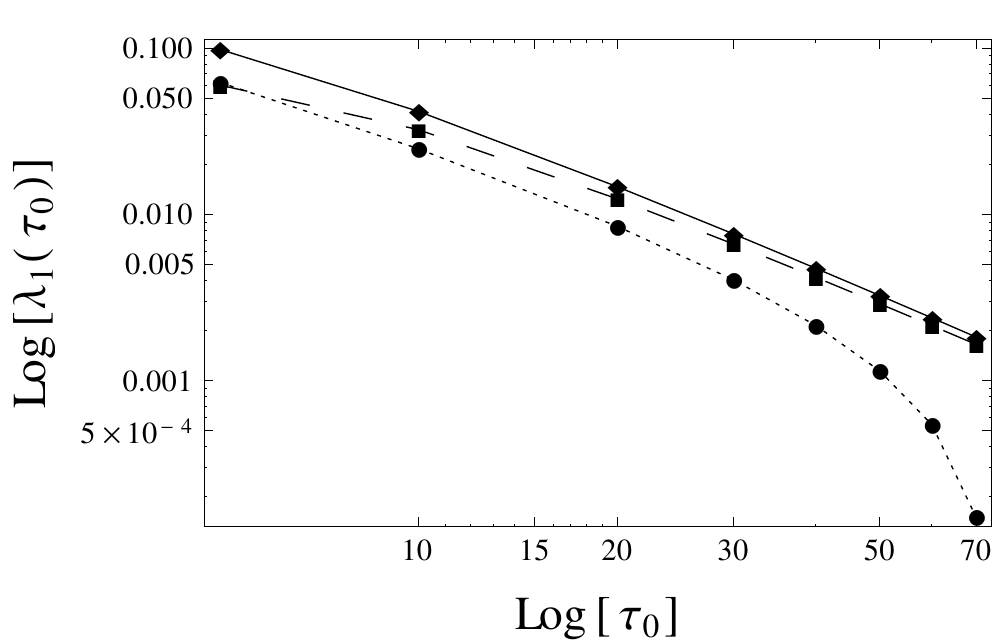}
  \caption{Comparison between the first eigenvalue of the space operator $\mathcal L_\tau$ defined in eq.~\rf{Loperator} obtained with different methods as a function of the optical depth $\tau_0$ of the slab. Filled circles: Atkinson-Shampine method (Sect. \ref{ASmethod}). Filled squares: variational approach  by L88 (see eq.~\ref{lambda:var}). Filled rhombuses: asymptotic limit as defined in equation \rf{lambda:Fourier}.} 
  \label{fig:eigenv_B} 
\end{figure}
\noindent
In Fig.~\ref{fig:eigenv_B}, we compare the first eigenvalue that we obtain from numerical computations with respect to the analytical estimates performed in L88. 
Lyubarskii provides two estimates of the first eigenvalue: one is obtained performing a Fourier transform of the kernel (\ref{kerneltheta}), assuming $\tau_0\rightarrow\infty$, which gives
\begin{equation}
\lambda_F=\frac{\pi^2}{4\tau_0^2}(\log\,4\tau_0-2), \label{lambda:Fourier} 
\end{equation}
instead, the other estimate, which is 
\begin{equation}
\lambda_{VAR}=\frac{5}{2\tau_0^2}\left(\log\,8\tau_0+\gamma-\frac{13}{3}\right), \label{lambda:var}
\end{equation}
is found solving the equation (\ref{spatialeqn}) with a variational method (here $\gamma$ is the Euler's constant). 
As suggested by the author, the relation (\ref{lambda:Fourier}) is no longer satisfied if we are in the case of large optical depths ($\tau_0\gtrsim20$). Nevertheless, the eigenvalue obtained through equation (\ref{lambda:var}) is, by definition, an upper limit of the exact value of $\lambda_1$, hence we expect smaller first eigenvalues for fixed optical depth.\\
At small $\tau$, the numerical computation and the two analytical estimates are quite similar, but when the optical depth increases, the numerical eigenvalues begin to deviates significantly from the analytical estimates.

\begin{figure}[t!]
  \centering
  \includegraphics[scale=.8]{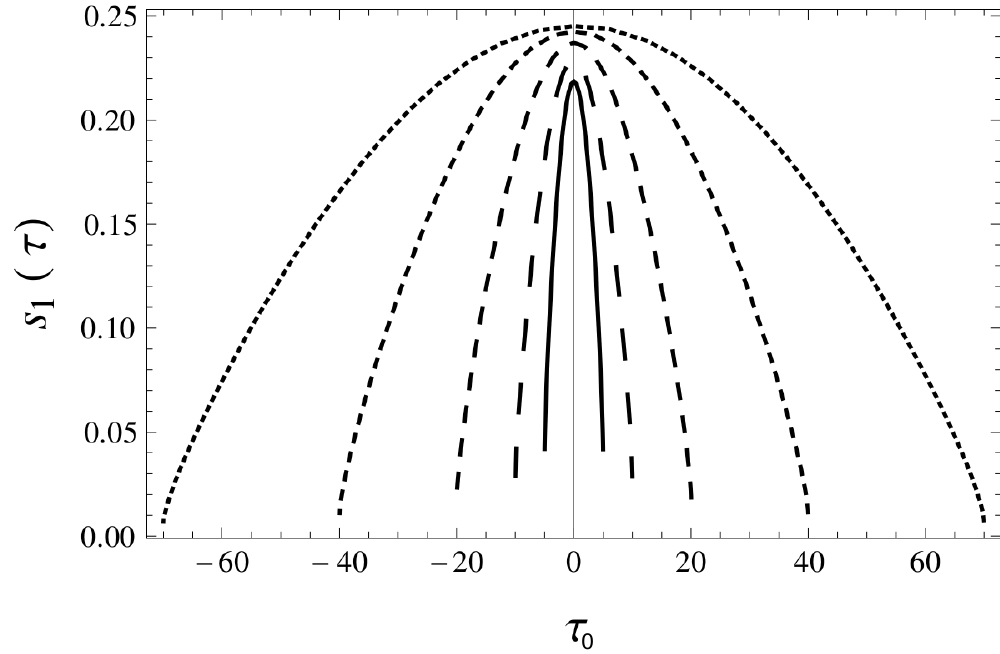}\qquad
  \caption{First eigenfunctions of the space operator $\mathcal L_\tau$ defined in eq.~\rf{Loperator} for different values of maximum optical depth $\tau_0=5,10,20,40,70$ (from solid to dotted line as $\tau_0$ increases). 
  }
  \label{fig:boundary} 
\end{figure}
\noindent
The left panel of Fig.~\ref{fig:boundary} represents the first eigenfunction for different values of maximum optical depth $\tau_0$. The plot shows that, in analogy with other physical situations, like, for instance, a potential well with increasing height, eigenfunctions are decreasing at the boundaries, approaching zero as $\tau\rightarrow\infty$. \\
Then, we suggest that the larger is the optical depth the smaller is the number of photons which escape from the slab boundary, the closer is the regime of saturated Comptonization. The result is compatible with the analytic expression of the first eigenfunction considered in L88 (see eq.~28 of L88II) in which the integral term gives a nonzero contribution at $\tau=\tau_0$.

\begin{figure}[t!]
  \centering
  \includegraphics[scale=.8]{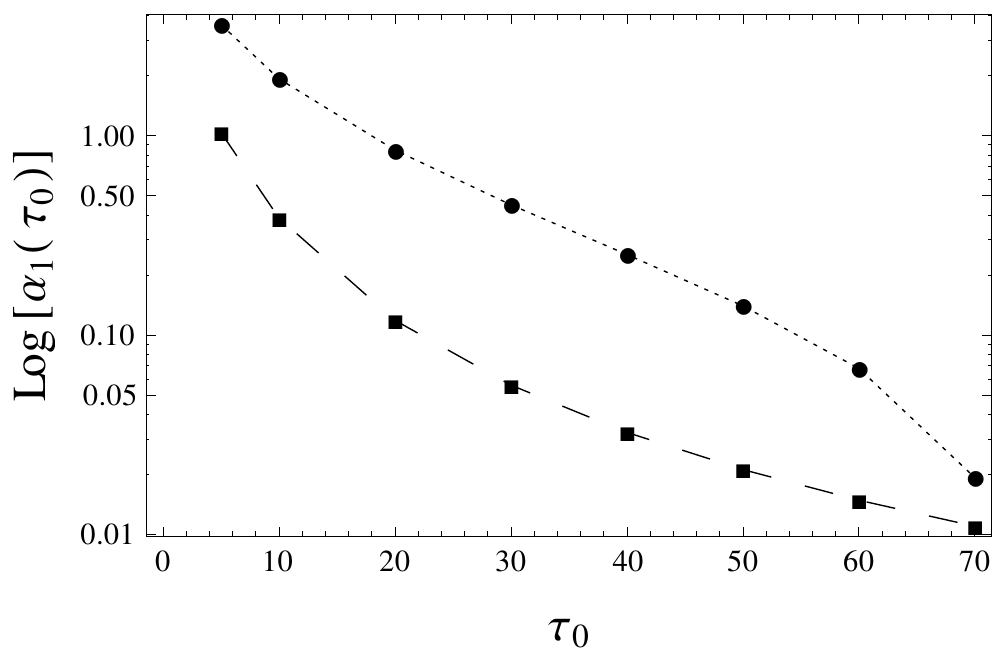}
  \caption{Spectral indexes of the Comptonization spectrum derived from the first eigenvalue  of the space problem \rf{spatialeqn} in the presence of a strong magnetic field ({\it filled circles})  and for the unmagnetized case   ({\it filled squares}). 
The explicit forms of $\alpha_1(\tau_0)$ are given in equations (~\ref{specindex}) and \rf{alphanomag}, respectively. Note that
for a fixed optical depth $\tau_0$ of the slab, the presence of the magnetic field, through reduction of the scattering cross-section, gives rise to softer spectra than an unmagnetized plasma.}  
  \label{fig:index}
\end{figure}
\noindent
In Fig.~\ref{fig:index} we compare the spectral index of the emerging spectrum for increasing maximum optical depth $\tau_0$ for a strongly magnetized system with the case of unmagnetized plasma. The analytic expression of the spectral index in the former case is presented in equation (\ref{specindex}). 
Following \cite{SunyaevTitarchuk:1980}, hereafter ST80, the spectral index for a slab/disk geometry in absence of magnetic field is described by the relation
\begin{equation}
 \alpha_{\rm{UNMAG}}=-\frac{3}{2}+\sqrt{\frac{9}{4}+\frac{m_e c^2}{kT_{\rm e}}\lambda}, \label{alphanomag}
\end{equation}
where 
\begin{equation}
 \lambda=\frac{\pi^2}{12\left(\tau_0+\frac{2}{3}\right)^2}. \label{1lambdanomag}
\end{equation}
Also in this case, this eigenvalue represents the leading term in the series \rf{solution} and mostly dictates the shape of the emerging spectrum. We find that the spectral index of the first Comptonization order in the strong magnetic field case is larger with respect to the unmagnetized case for any value of optical depth. \\
Indeed, the magnetic field makes the Comptonization process less efficient overall ($y_{\rm{NR}}^{\rm{mag}}\sim0.13\,y_{\rm{NR}}^{\rm{nomag}}$), as a consequence of the reduction of the scattering cross-section (see eq.~\ref{crosssec2}) for photons travelling at right angles with respect to the magnetic field direction.

\begin{figure}[t!]
  \centering
  \includegraphics[scale=.8]{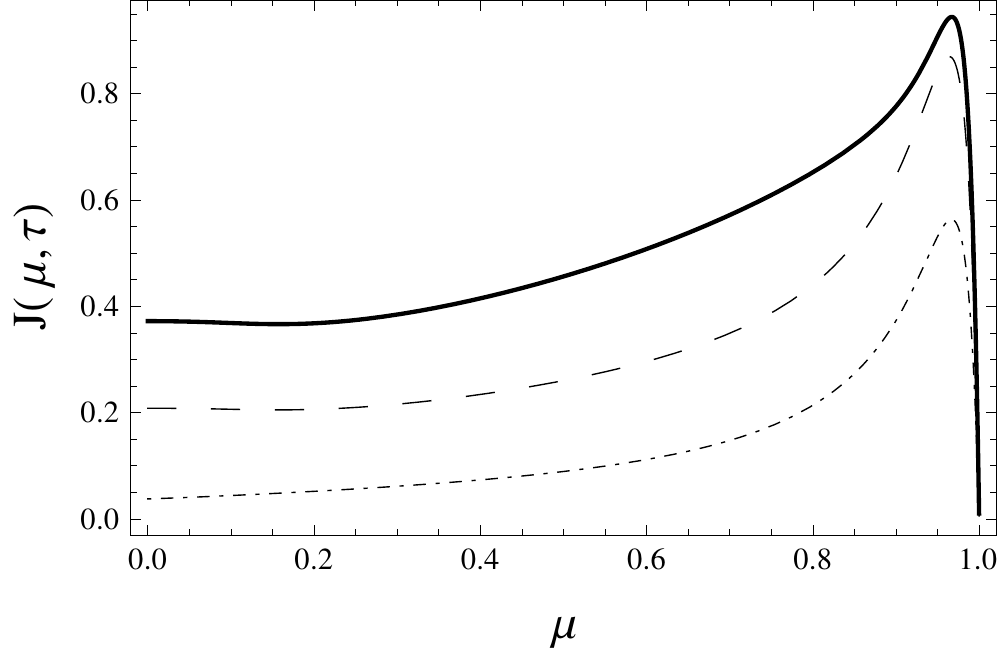}
  \includegraphics[scale=.8]{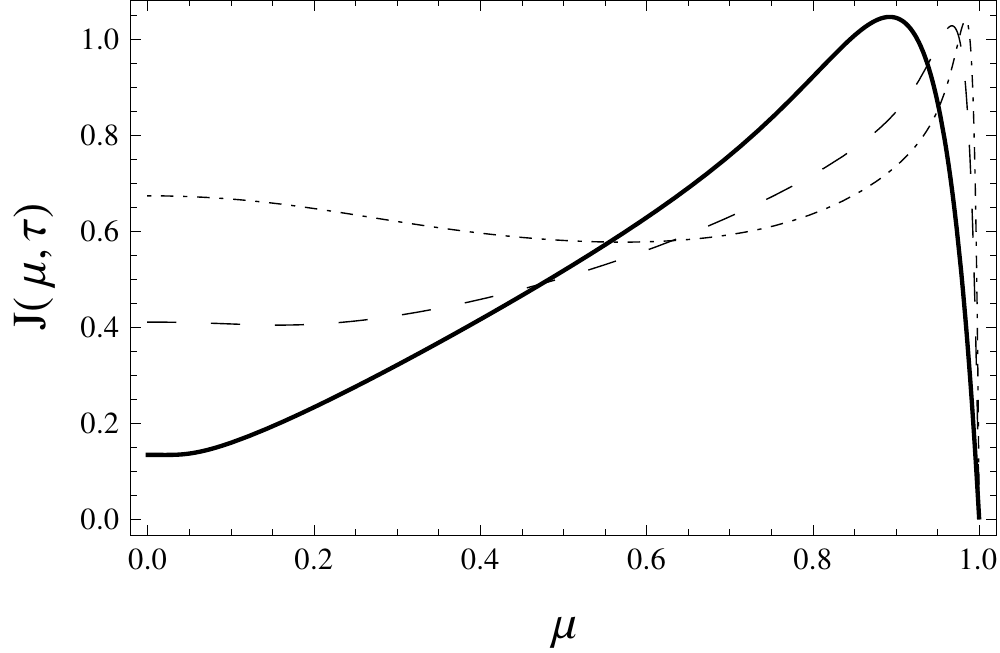}
  \caption{\emph{Left Panel}: Angular distribution of ordinary photons for $0\leq\mu\leq1$ and optical depth $\tau_0=20$. Solid line: angular distribution for the sum of the first 10 eigenfunctions and relative coefficients $c_k$ obtained with the algorithm described by AS08. Dashed line: same function calculated with the first eigenfunction and coefficient from the algorithm of AS08. Dot-dashed line: same function but defined as in equation (29) of L88. \emph{Right panel}: Angular distribution of the ordinary photons emerging at the top of the slab $(0<\mu<1)$ for eigenfunctions and relative weights up to $k=10$ for the cases of optical depth $\tau_0=5$ (solid line), $\tau_0=20$ (dashed line) and $\tau_0=40$ (dot-dashed line).}
  \label{fig:J} 
\end{figure}
 
\noindent
Unlike the case of the radiative transfer problem reported in ST80, the presence of the magnetic field induces a significant angular dependence in the specific intensity, which is taken into account by the function $J(\tau,\mu)$ defined in equation \rf{bc_ang}.\\
Considering the case of the dominant Comptonization mode k=1, on the left panel of Fig.~\ref{fig:J} we compare the angular distribution obtained with the AS08 algorithm described in Sect.~\ref{ASmethod} considering the sum of the first 10 eigenfunctions and relative coefficients and the same distribution but with the leading term only and the angular distribution calculated using the first eigenfunction corrected at the boundaries (eq.~29 in L88). \\
The distributions appear qualitatively the same, except for a scale factor. The normalization gap between the L88 estimate and the numerical estimate with the first eigenfunction is due to the coefficient $c_1$ multiplying the latter. Basically the same gap is introduced between the two numerical estimates by considering the sum of the first 10 terms of the series \rf{solution} in \rf{bc_ang}, although the shape remains substantially unchanged. Therefore, terms with $k>1$ contributes to the angular photon distribution as a scale factor.\\
The right panel in Fig.~\ref{fig:J} presents the change in the angular distribution for $\tau_0=5,20,40$. The peak of the distribution becomes narrow and is moving towards $\mu=1$, where the function goes to zero, as the optical depth increases. As expected, for $\tau\rightarrow\infty$ the function $J$ tends to be flatter, except for $\mu\sim1$ where the peak is located, approaching to an isotropic distribution of the photons.\\    
As for increasing values of optical depth, the probability for photons to escape the plasma progressively decreases, the eigenfunctions tend to vanish at the boundaries of the slab, meaning that only a small amount of photons find their way out through the plasma. The angular distribution approaches a flat, isotropic distribution and the system is entering the regime of saturated Comptonization.

\begin{table}[t!]
\centering
\begin{tabular}{rcc}
              \toprule %
              \toprule %
              $k$        & $\lambda_k(\tau=\tau_0)$ & $\alpha_k(\tau=\tau_0)$\\\toprule\toprule%
                                              1    & 0.0085  & 0.8460 \\
                                              2    & 0.0297  & 2.1922 \\
                                              3    & 0.0555  & 3.3498 \\
                                              4    & 0.0838  & 4.3623 \\
                                              5    & 0.1130  & 5.2496 \\\bottomrule\bottomrule
\end{tabular}
\caption{First five eigenvalues $\lambda_k$ of the space operator \rf{Loperator} and derived spectral index $\alpha_k$ (see eq.~ \rf{specindex} for a slab with optical depth $\tau_0=20$. }
\label{tab:lambda_alpha_k}
\end{table}

\begin{table}[t!]
\centering
\begin{tabular}{rcc}
	      \toprule %
	      \toprule %
	      $\tau_0$        & $\lambda_1(\tau_0)$ & $\alpha_1(\tau_0)$\\\toprule\toprule%
					      5    & 0.0622  & 3.6101 \\
					      10   & 0.0248  & 1.9290 \\
					      20   & 0.0085  & 0.8460 \\
					      30   & 0.0040  & 0.4499 \\
                                              40   & 0.0021  & 0.2533 \\
					      50   & 0.0011  & 0.1399 \\
					      60   & 0.0005  & 0.0680 \\
                                              70   & 0.0001  & 0.0193 \\\bottomrule\bottomrule
\end{tabular}
\caption{Same of Tab.~\rf{tab:lambda_alpha_k} but for different values of the optical depth $\tau_0$ of the slab.}
\label{tab:lambda_alpha_tau}
\end{table}

\subsection{Energy equation and specific intensity}
\label{energy}

\begin{figure}[t!]
\centering
  \includegraphics[scale=.8]{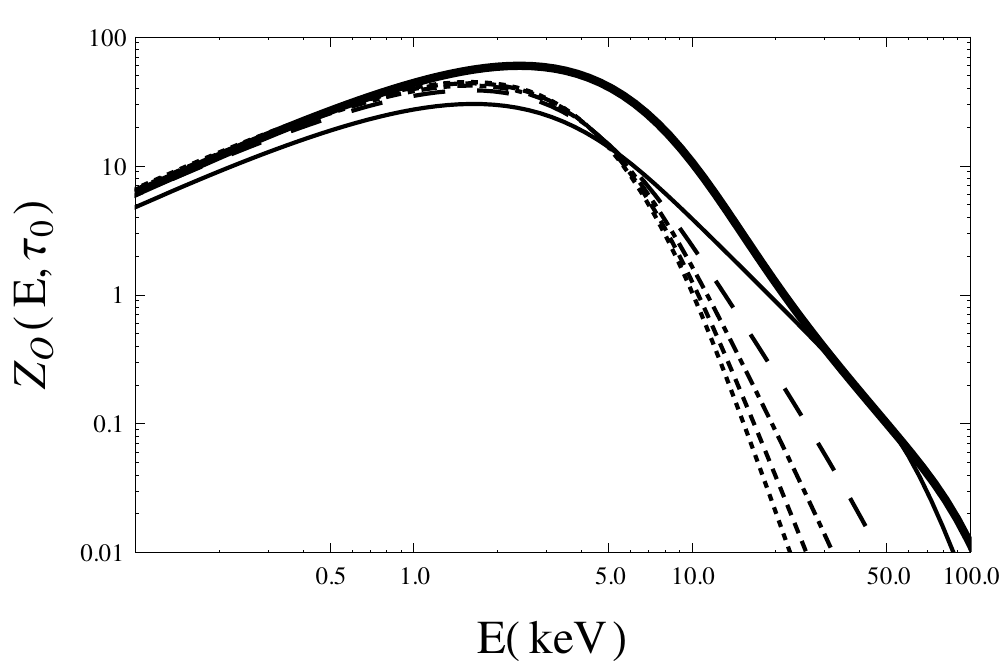}
  \includegraphics[scale=.8]{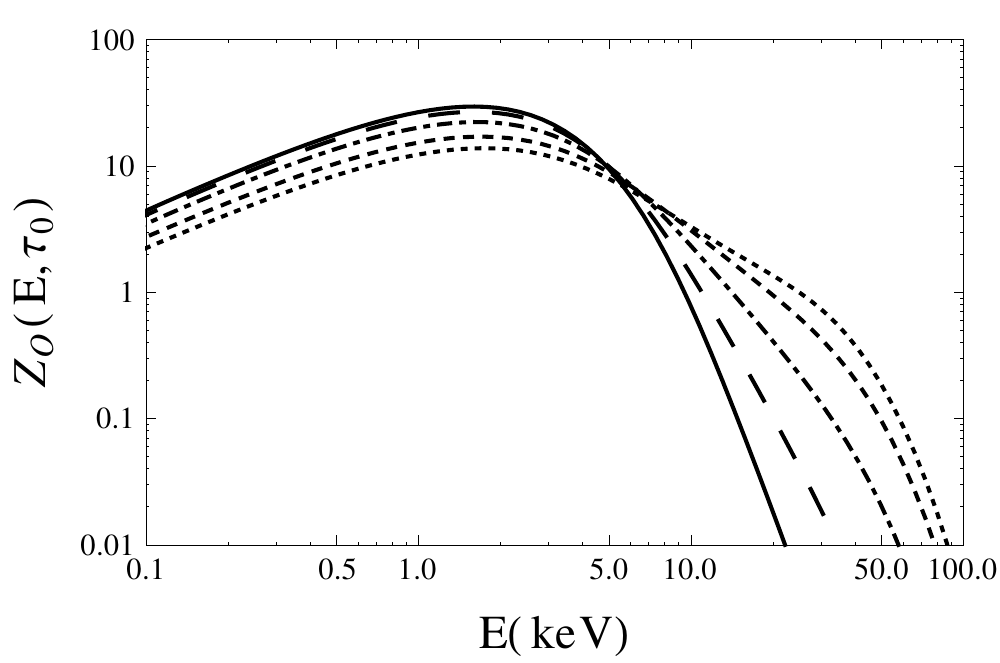}
  \caption{\emph{Left panel}: Emergent spectra in arbitrary units, solution of equation \rf{energyeqn} for eigenvalues $\lambda_k$ with $k=1,2,3,4,5$ (see Tab.~\rf{tab:lambda_alpha_k}), maximum optical depth $\tau_0=20$, blackbody temperature $kT_{bb}=1$ keV and electron temperature $kT_{\rm e}=10$ keV. Thin solid line: energy flux for $k=1$. Large dashed line: $k=2$. Dot-dashed line: $k=3$. Small dashed line: $k=4$. Dotted line: $k=5$. Thick solid line: total emergent spectra obtained by the sum of the first five terms of the series weighted with the corresponding expansion coefficients $c_k$. The total emergent spectrum has been multiplied by an arbitrary scaling factor in order to be compared with its components. \emph{Right panel}: Emergent spectra in arbitrary units, solution of equation \rf{energyeqn}, for spectral index $\alpha(\tau,\lambda_1)$ defined in the relation \rf{specindex} and increasing maximum optical depth with $kT_{bb}$ and $kT_{\rm e}$ as specified on the left panel. Solid line: $\tau_0=5$. Large dashed line: $\tau_0=10$. Dot-dashed line: $\tau_0=20$. Small dashed line: $\tau_0=40$. Dotted line: $\tau_0=70$. The values of the spectral indexes and eigenvalues are reported in Tab.~\ref{tab:lambda_alpha_tau}}  
  \label{fig:fluxes}
\end{figure}

In order to find solutions of the energy operator defined in equation \rf{energyeqn}, we need to specify a source term on the right-hand side. Here we consider the case of blackbody seed photons with an exponetially attenuated spatial distribution, described as
\begin{equation}
 S(x, \tau)=C e^{-\tau/2\tau_0}kT_{\rm e}^3\frac{x^3}{e^{x(kT_{\rm e}/kT_{\rm{bb}})}-1}
\end{equation}
where $kT_{\rm{bb}}$ and $kT_{\rm e}$ are the photons and the electron temperatures, respectively. The constant $C$ is a normalization depending on the specific problem. \\
The energy-dependent part of the specific intensity is obtained through the convolution with the Green's function of the energy operator (equations \ref{startingpoint}-\ref{flux2}).
On the left panel of Fig.~\ref{fig:fluxes}, we show results for the first five term of the series defined in equation \rf{flux2}, solution of the problem, i.e. for eigenvalues $\lambda_k$ with ($k=1,2,3,4,5$) for the case of $\tau_0=20$, blackbody temperature $kT_{bb}=1$ keV and electron temperature $kT_{\rm e}=10$ keV. 
As expected, the high-energy part of the spectrum is mostly determined by the first Comptonization mode, following the qualitatively similar behaviour of the unmagnetized case (see ST80). \\
On Tab.~\ref{tab:lambda_alpha_k}, we present the eigenvalues $\lambda_k(\tau=\tau_0)$, found as described in the Par.~\ref{space}, and the spectral index $\alpha_k$ for $k=1,2,3,4,5$ and for $\tau_0=20$. As we expect from relation \rf{specindex}, the larger is the eigenvalue, the larger is the index, which means that the spectrum becomes steeper and steeper. \\
In particular, the step between the first and the second eigenvalue is peculiar: even though eigenvalues and indexes are, in good approximation, equally separated, basically the first term provides the most evident deviation from the seed spectrum.\\ From a phenomenological point of view, the total spectrum can be split into two components: the first term $k=1$ of the series \rf{solution} represents the efficiently Comptonized seed photon, while the remaining terms $(k\geqslant 2)$ describe the photons emerging from the bounded medium without appreciable modification of their energy. The relative contribution of the two features eventually depends on the spatial distribution of the seed photons (e.g. TMK97; \citealp{TitarchukZannias:1998, LaurentTitarchuk:1999}).\\
On the right panel of Fig.~\ref{fig:fluxes}, we present instead solutions relative to mode $k=1$ for increasing optical depth $\tau_0=5,10,20,40,70$ and $kT_{bb}=1$ keV and $kT_{\rm e}=10$ keV. Varying $\tau$ corresponds to a change in the spectral index $\alpha$ as pointed out on Tab.~\ref{tab:lambda_alpha_tau}. \\
Of course, the larger is the optical depth, the flatter becomes the spectrum. Smaller spectral indexes $\alpha$ imply substantially Comptonized spectra for $\tau\rightarrow\infty$ up to the asymptotic regime of saturated Comptonization.  
\begin{figure}[t!]
\centering
  \includegraphics[scale=.9]{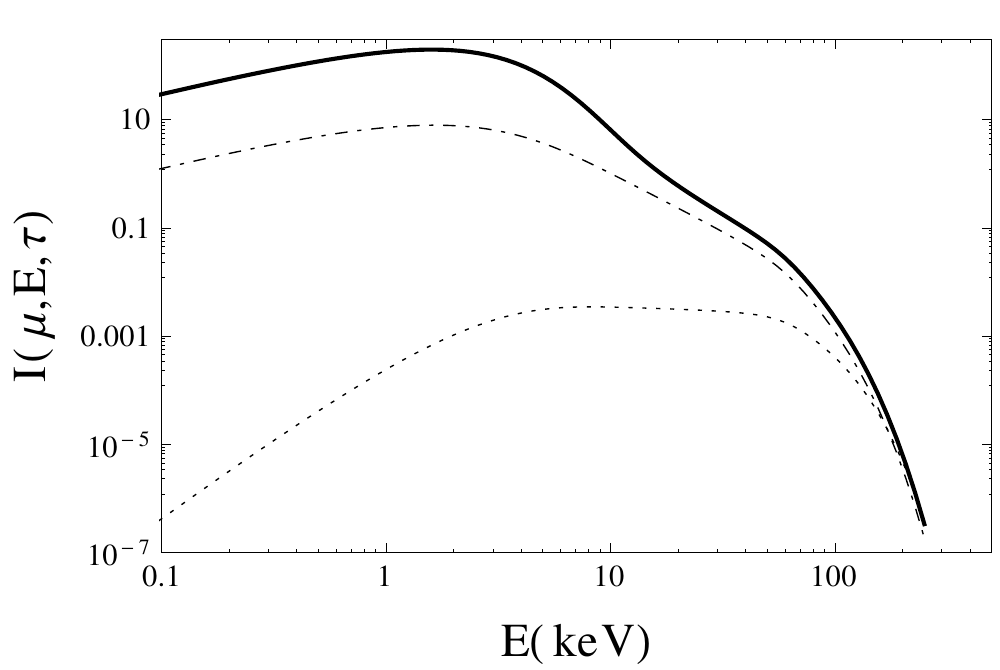}
\caption{Solid line: total specific intensity of ordinary photons obtained by the sum of the first 10 terms of the Fourier series of equation  \rf{sint}.  Dot-dashed line: specific intensity of ordinary photons for the first Comptonization model $k=1$ (eq. \ref{sintO}). Dashed line: specific intensity of extraordinary photons created via mode-switching from a fraction of ordinary photons as calculated in equation \rf{sintE}. The optical depth of the slab is $\tau_0=20$, the seed photon spectrum is a blackbody with  $kT_{bb}=1$ keV while the electron temperature is $kT_{\rm e}=10$ keV. The case of photon escaping with an angle $\mu=0.5$ with respect to the slab normal is shown. }
  \label{fig:sint}
\end{figure}
In order to concentrate on the effectively Comptonized spectrum, we consider the first term of the series of equation \rf{sint} and write the specific intensity for the ordinary photons as
\be
I(x,\mu,\tau)\approx c_1J_1(\mu,\tau_0)\mathcal Z_1(x) \label{sintO}
\ee
where the angular distribution $J_1(\mu,\tau_0)$ is calculated in equation \rf{bc_ang}, $\mathcal Z_1(x)$ is given by the
convolution of equation (\ref{flux2}) relative to  the first eigenvalue $\lambda_1(\tau)$ and $c_1$ is the first coefficient
of the Fourier series obtained from the projection of the seed photon space distribution over the first space eigenfunction.\\
We report in Fig.~\ref{fig:sint} the  specific intensity for the ordinary photons that keep their polarization during the scattering process, by considering the first 10 terms of the Fourier series  defined in equation \rf{sint} for a given parameter
set of seed photons blackbody temperature, optical depth and temperature of the electrons of the slab, and
emerging angle of the radiation field with respect to the slab normal.  The latter variable is mild and does not affect the 
spectral shape.
Additionally, we also show the ordinary-photon specific intensity relative to the first Comptonization mode only (k=1) and the specific intensity of the fraction of ordinary photons which become instead extraordinary photons. The latter feature  has been calculated through the relation \rf{sintE} and has been assumed isotropically distributed since the cross-section \rf{crosssec21} doesn't depend on the angle of propagation of the scattered photons. 

\section{Discussion}
\label{discussion}

The numerical method developed in this paper to investigate the spectral formation for photons propagating in a slab dipped into a strong magnetic field allows us to perform some qualitative and quantitative considerations about the main spectral features of the emerging spectra.\\
Following L88, we have calculated the spectrum of the subset of the initial seed photon population that has the largest probability to be Comptonized, based on the cross-sections \rf{crosssec2}-\rf{crosssec12}.
Considering photons propagating at large angles to the field and keeping in mind that we assume photon energies well below the cyclotron energy ($h\nu_g\sim 10$ MeV for $B\sim 10^{14}$ G), the ordinary photons are those which have more chances to undergo a considerable amount of scatterings. Moreover, the presence of the cross-section \rf{crosssec21} in eq.~\rf{lyubeqn} tells us that a fraction of the ordinary photons will change their polarization turning into extraordinary photons. \\
Separating the variables, we were able to solve two independent equations (\ref{spatialeqn})-(\ref{energyeqn}). \\
The equation for the space operator \rf{Loperator} is an integral eigenvalue problem, that we solved numerically developing the technique described in AS07.
The eigenfunctions of the space operator that we obtain behave like eigenfunctions of many other physical systems, like a potential well. Indeed, as the optical depth increases the eigenfunctions approach zero at the boundaries, meaning that photons remain trapped into the medium and undergo a large amount of scatterings. 
The angular photon distribution calculated from the eigenfunctions of the space equation \rf{spatialeqn} reflects the same behaviour. For progressively increasing values of the optical depth of the system, the angular distribution flattens and the photon field becomes almost isotropized.\\
The emerging energy spectrum, solution of the system (\ref{spatialeqn})-(\ref{energyeqn}), can be essentially split into two main components. The first is a soft peak, given by the photons which escaped the system without appreciable scatterings, and represented by the Comptonization modes $(k\geqslant 2)$  of the expansion series. The second term, corresponding to $(k=1)$ is the actual Comptonized spectrum whose spectral slope depends on the total optical depth.
For increasing optical depth the spectrum becomes flatter, approaching the regime of saturated Comptonization when $\tau_0\rightarrow\infty$. Even though the spectral shape and the dependence on the eigenvalues is quite similar to the unmagnetized case, the comparison between the spectral indexes reveals that for fixed physical set-up, such as electron temperature and optical depth, Comptonization is less efficient if a strong magnetic field is present.  \\ 
In the calculation of the total spectrum we took into account also the contribution of the extraordinary photons created via mode switching. However, it provides a small, or even negligible, contribution to the overall specific intensity calculated here, as we saw in Fig.~\ref{fig:sint}. Nonetheless, the contribution provided by the extraordinary photon's specific intensity is expected to be completely Comptonized, since it originates from photons which have already scattered several times before the flip in polarization. \\
It is worth pointing out that the so-obtained extraordinary mode spectrum does not necessarily represents the total contribution of the extraordinary photons to the emerging spectrum. It is reasonable to guess, though, that the total extraordinary spectrum should emerge from the slab basically unchanged with respect to their initial distribution, namely a Planck spectrum. \\
The dominance between the two total contributions, ordinary and extraordinary, can be established by several factors, like for instance the initial relative percentage of the two seed photon populations and the truncation of the series in eq.~\rf{sint}. Even if we assume that both contributions have the same weight on the formation of the total spectrum, distinguishing between the two from the point of view of data analysis is highly unlikely, especially in the soft thermal peak region.  \\
Therefore, even if solving the coupled-system of radiative transfer equations \citep{Lyubarky:2002} is the best approach for tackling the problem of radiative transfer in strong magnetic fields and describing the propagation and the mutual interaction of the two photons' polarization mode, the advantage in the spectral fitting will not be so pronounced, since the two contributions can be hardly disentangled.\\


\section{Conclusions}
\label{Conclusions}
We have performed a numerical study of the radiative transfer problem for a plane-parallel slab dipped into a strong magnetic field, focusing on the ordinary-mode photons, which have the less-suppressed scattering cross section for energies below the cyclotron energy, which of order of 10 MeV for a magnetic field of order of $10^{14}$ G. 
The full solution of the RTE is obtained by the Fourier's method and can be described as a series of terms in which the dependencies on the independent variables (here energy and optical depth) of the problem are decoupled. 
The simple angular dependence of the Thompson cross section, out of the Klein-Nishina regime, restricts the range of physical applicability of the derived spectra up to energies $\sim 100$ keV.\\
Our work can be considered as a completion and check, using numerical techniques, of the analytical results reported by L88.
A thorough and careful treatment of the singularity  of the kernel characterizing the space diffusion operator  allowed us to compute the series of eigenvalues and eigenfunctions of the coupled energy-space problem. 
Actually, the implemented numerical methods, would in principle give the possibility to find as many as desired terms of the Fourier series reported in equation (\ref{solution}), albeit the ordinary-mode Comptonized spectra are dominated by the first term. We have compared our solution with the analytical estimate given by L88 and we have also performed a comparison of the spectral index of the Comptonization spectrum  with
the case of unmagnetized plasma.\\
The geometrical configuration here considered for our computation, namely a simple static slab with a magnetic field parallel to its normal, makes difficult a straightforward application of the model to astrophysical objects at all, because of the complex shape of the magnetic field in magnetar or X-ray pulsars out of the neutron star surface (e.g. multipole components).\\
However, in the study of the spectral formation close to the polar caps of a neutron star dominated by a dipole component, our assumptions could work in good approximation allowing to treat the angular dependence of the emerging spectra, at least for ordinary-photons, instead of using the zero-moment approximation for the radiation field.

\section{Acknowledgement}

The authors would like to thank Alessandro Drago for the very helpful discussions and Roberto Turolla for the critical feedback provided in occasion of Ceccobello's PhD thesis. 


\bibliographystyle{natbib}
\bibliography{mybiblio}

\end{document}